\documentclass[reprint,amsmath,amssymb,aps,prl]{revtex4-2}

\usepackage{comment}
\usepackage{enumitem}
\setlist[enumerate]{itemsep=0mm}
\usepackage{dcolumn}
\usepackage{bm}
\usepackage{color}
\usepackage{graphicx}
\usepackage{hyperref}
\usepackage{braket}
\usepackage{bbm}
\usepackage{tikz}
\usepackage{tikz-3dplot}
\usepackage[outline]{contour}

\newcommand{\be}{\begin{equation}}
\newcommand{\ee}{\end{equation}}
\newcommand{\ba}{\begin{aligned}}
\newcommand{\ea}{\end{aligned}}
\newcommand{\1}{\mathbbm{1}}

\begin{document}

\title{
Macroscopic effects of localised measurements\\in jammed states
of quantum spin chains}

\author{Kemal Bidzhiev}
\affiliation{%
 Universit\'e Paris-Saclay, CNRS, LPTMS, 91405, Orsay, France
}%
\author{Maurizio Fagotti}%
 \email{maurizio.fagotti@universite-paris-saclay.fr}
\affiliation{%
 Universit\'e Paris-Saclay, CNRS, LPTMS, 91405, Orsay, France
}%
\author{Lenart Zadnik}%
\affiliation{%
 Universit\'e Paris-Saclay, CNRS, LPTMS, 91405, Orsay, France
}%

\begin{abstract}
A quantum jammed state can be seen as a state where the phase space available to particles shrinks to zero, an interpretation quite accurate in integrable systems, where stable quasiparticles scatter elastically.
We consider the integrable dual folded XXZ model, 
which is equivalent to the XXZ model in the limit of large anisotropy.  We perform a jamming-breaking localised measurement in a jammed state. We find that jamming is locally restored, but local observables  
exhibit nontrivial time evolution on macroscopic, ballistic scales, without ever relaxing back to their initial values.
\end{abstract}

\maketitle

\paragraph{Introduction.} 
Recent remarkable progress in quantum information science owes much to the advancement of cold-atom and trapped-ion setups~\cite{JAKSCH2005,Bloch2012,Barreiro2011,Tan2015}. 
These have become so refined~\cite{Rydberg_simulation2,Zhang2017,Supremacy,QuantumSimulators256} to allow for the design of so-called ``quantum simulators''~\cite{QuantumSimulators}. Understanding the effects of perturbations and measurements has always been crucial in this regard, be it due to the role of perturbations in the decoherence process, or the one of measurements in simulation protocols~\cite{SCHLOSSHAUER2019}. Integrable~\cite{Polkovnikov2017,zadnik2018,zadnik2019,lucas2021} and generic~\cite{Bertini2018,Nahum2017,chan2018,keyserlingk2018,khemani2018,bensa2021} quantum circuits are potentially useful in this endeavour: they allow for some degree of exact treatment~\cite{aleiner2021,claeys2021,sarang18,alba2019,klobas2021} and befit experimental realisation~\cite{potocnik2015,Neill2021}.

Localised measurements can be viewed as a type of ``quantum quench'', i.e., the non-equilibrium dynamics induced by a sudden perturbation, studied in the last decades especially in the context of relaxation in quantum many-body systems (see~\cite{Calabrese2016} and articles therein). Such perturbations break the homogeneity of the system, making its study challenging.  
In integrable models the most effective large-scale description of the dynamics in the presence of inhomogeneities is arguably the so-called  ``generalised hydrodynamics'' (GHD)~\cite{LQSS0,ben,bruno}. Although GHD correctly predicts the large-scale dynamics of local observables in numerous quench protocols (e.g., in the two-temperatures scenario), the information provided by the theory is sometimes incomplete.
The first example of this kind was exhibited in Ref.~\cite{piroli2017transport}, considering the massive Heisenberg model: the ingredients of GHD are blind to observables that are odd under spin flip, entailing the inclusion of an additional independent continuity equation. An even more striking example was considered in Refs~\cite{eisler2020front,gruber2021entanglement,zauner2015time,eisler2016universal,eisler2018hydrodynamical}, in which GHD keeps a symmetry that is instead broken in the thermodynamic limit: observables not respecting that symmetry are affected by a class of localised perturbations at arbitrarily long times and large distances from the inhomogeneity.

\begin{figure}[ht!]
    \centering
    \begin{tikzpicture}[scale=0.9]
    \draw[fill=white, rounded corners = 2,thick] (-0.35,-1.35) rectangle ++(3.7,4.1);
    \foreach \y in {1,...,2}
        \draw[black] (-3-0.1,0.25*\y) -- (3+0.1,0.25*\y);
    \foreach \y in {6,...,7}
        \draw[black] (-3-0.1,0.25*\y) -- (3+0.1,0.25*\y);
    \foreach \y in {1,...,2}
        \draw (-3-0.17,0.25*\y) circle (1.5pt);
    \foreach \y in {6,...,7}
        \draw (-3-0.17,0.25*\y) circle (1.5pt);
    \foreach \y in {1,...,2}
        \draw (3+0.17,0.25*\y) circle (1.5pt);
    \foreach \y in {6,...,7}
        \draw (3+0.17,0.25*\y) circle (1.5pt);
    \draw[fill=green!10!white!95!black,opacity=1,rounded corners=2] (-2.25,-0.5) rectangle ++(1.75,3) node[black,midway]{$e^{i H t}$};
    \draw[fill=green!10!white!95!black,opacity=1,rounded corners=2] (2.25,-0.5) rectangle ++(-1.75,3) node[black,midway]{$e^{-i H t}$};
    \fill[white] (-2.3,2.25) rectangle ++(1.85,0.4);
    \fill[white] (2.3,2.25) rectangle ++(-1.85,0.4);
    \fill[white] (-2.3,-0.65) rectangle ++(1.85,0.4);
    \fill[white] (2.3,-0.65) rectangle ++(-1.85,0.4);
    \node[anchor=center] at (-2.75,1.115) {$\vdots$};
    \node[anchor=center] at (2.75,1.115) {$\vdots$};
    \node[anchor=center] at (0,2.5) {$\vdots$};
    \node[anchor=center] at (0,-0.25) {$\vdots$};
    \node[anchor=center] at (-2.75,-0.25) {$\vdots$};
    \node[anchor=center] at (2.75,-0.25) {$\vdots$};
    \draw[fill=white,opacity=1,rounded corners=0,thick] (-2.95,1.6) rectangle ++(0.5,0.3);
    \draw[fill=white,opacity=1,rounded corners=0,thick] (-0.175,0.325) rectangle ++(0.35,0.35) node[black,midway]{\scriptsize{$W$}};
    \draw[fill=white,opacity=1,rounded corners=0,thick] (2.95,1.6) rectangle ++(-0.5,0.3);
    \node[anchor=center] at (-2.7,2.125) {\scriptsize{$O$}};
    \node[anchor=center] at (2.7,2.125) {\scriptsize{$O$}};
    \draw[black,line width=0.6pt,->] (-2.8+0.075,1.65) -- (-2.6+0.075,1.85);
    \draw[black,line width=0.6pt,->] (2.6+0.075,1.65) -- (2.8+0.075,1.85);
    \draw[black,line width=0.6pt] (-2.9,1.675) to[out=30, in=150] (-2.5,1.675);
    \draw[black,line width=0.6pt] (2.5,1.675) to[out=30, in=150] (2.9,1.675);
    \node[anchor=center] at (0,1) {\scriptsize{\textbf{Bob}}};
    \node[anchor=center] at (-2.85,2.375) {\scriptsize{\textbf{Alice}}};
    \node[anchor=center] at (2.85,2.375) {\scriptsize{\textbf{Alice}}};
    \fill[black,opacity=0.075] (-3.5,-1.35) rectangle ++(3.05,4.1);
    \node[anchor=center] at (-2.75,-1.1) {\scriptsize{$(1)$}};
    \node[anchor=center] at (-1.375,-1.1) {\scriptsize{$(2)$}};
    \node[anchor=center] at (0,-1.1) {\scriptsize{$(3)$}};
    \node[anchor=center] at (1.375,-1.1) {\scriptsize{$(4)$}};
    \node[anchor=center] at (2.75,-1.1) {\scriptsize{$(5)$}};
    \end{tikzpicture}
    \caption{The ``wing-flap'' protocol: (1) Alice measures an observable $O$; (2) the system evolves for a time $-t$; (3) Bob applies a ``wing-flap'' perturbation $W$; (4) the system evolves for a time $t$; (5) Alice measures $O$ again. 
    In Ref.~\cite{yan2020recovery} $W$ is a ``blind'' local projective quantum measurement at the position of $O$ (no information is kept), and step (5) is upgraded to a quantum state tomography. There it was shown that, if $H$ describes a generic system, the local state can be recovered  with a limited amount of effort.
    We simplify the protocol by trivialising the part highlighted in grey.}
    \label{fig:protocol_scheme}
\end{figure}
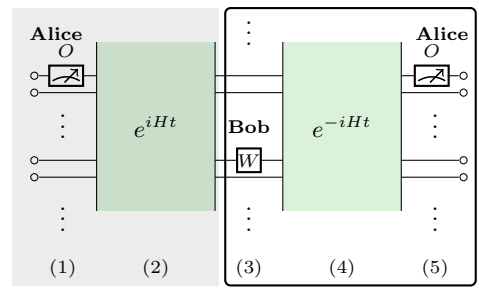

In this Letter we study the effect of a localised projective measurement in a quantum jammed state. To this end we consider a modification of the so-called ``wing-flap protocol'' (see Fig.~\ref{fig:protocol_scheme}) introduced in Ref.~\cite{campisi2017thermodynamics} to provide insight into the quantum information scrambling. The system is prepared in a low-entangled stationary state, which, we assume, is also an eigenstate of the operator measured by Alice. 
We then go back in time (a trivial step, since the state is stationary) considering an alternative history in which, at some unknown but fixed ancient time, Bob had performed an unknown projective quantum measurement at an unknown but fixed large distance from Alice.
Returning to an alternative present we wonder, following Ref.~\cite{yan2020recovery}, whether Alice can still recover the information she had without Bob's intervention.

Generally, in a shift invariant quantum spin chain the distant Bob's measurement is not expected to have any effect on Alice's subsystem. In particular, Ref.~\cite{gluza2019equilibration} showed that, in noninteracting spin chains, relaxation to a Generalised Gibbs ensemble (GGE) is not compromised by a localised perturbation. In our setting this implies that, at late times, the effect of Bob's measurement on a finite subsystem fades away. We are not aware of any physical argument against the generalisation of this result to interacting integrable systems. Indeed, this conclusion is supported by numerical investigations, which generally show the irrelevance of localised perturbations on the state after long-enough time. Consider, for example, the numerical tests of the generalised hydrodynamic predictions of time evolution after  two different states are joined together~\cite{Alba2021generalised}. The tests always differ in the way the states are joined, nevertheless the asymptotic behaviours at large times match. In generic systems scrambling is even more pronounced, so a very distant quantum measurement in a low-entangled stationary state is not expected to have visible effects however large the time is. Incidentally, even if the state were not stationary and Bob measured an observable in Alice's subsystem, Alice would  be able to recover the original local state~\cite{yan2020recovery}.

Here we apply this protocol to
the dual folded XXZ spin-$1/2$ chain, which belongs to a class of effective models emerging in strong coupling limits of spin chains~\cite{folded:1}. It corresponds to a special point of the two-component Bariev model~\cite{Bariev} and is described by the Hamiltonian
\be \label{eq:Hamiltonian}
H=J\sum_{\ell}(\sigma_\ell^x \sigma_{\ell+2}^x+ \sigma_\ell^y \sigma_{\ell+2}^y)\frac{\1- \sigma_{\ell+1}^z}{2}\, .
\ee
The model can be solved exactly by introducing two species of particles associated with spins up on either even or odd sites~\cite{folded:1} (see also~\cite{Pozsgay_folded_XXZ}). The solution is based on the observation that the unique nontrivial effect of $H$ corresponds to moving a spin up by two sites when it is adjacent to two spins down:
\begin{align}
\label{eq:process_pictures}
\begin{tikzpicture}[baseline=(current  bounding  box.center),scale=1]
\draw[black] (-3+0.1,0) -- (-2.5-0.1,0);
\draw[black] (-2.5+0.1,0) -- (-2-0.1,0);
\draw[black] (-2+0.1,0) -- (-1.5-0.1,0);
\draw[black] (-0.5+0.1,0) -- (0-0.1,0);
\draw[black] (0+0.1,0) -- (0.5-0.1,0);
\draw[black] (0.5+0.1,0) -- (1-0.1,0);
\draw[black] (2+0.1,0) -- (2.5-0.1,0);
\draw[black] (2.5+0.1,0) -- (3-0.1,0);
\draw[black] (3+0.1,0) -- (3.5-0.1,0);
\node[anchor=south] at (-2.75,0) {\contour{black}{$\uparrow$}};
\node[anchor=south] at (-2.25,0) {\contour{black}{$\downarrow$}};
\node[anchor=south] at (-1.75,0) {\contour{black}{$\downarrow$}};
\node[anchor=south] at (-0.25,0) {\contour{black}{$\uparrow$}};
\node[anchor=south] at (0.25,0) {\contour{black}{$\uparrow$}};
\node[anchor=south] at (0.75,0) {\contour{black}{$\downarrow$}};
\node[anchor=south] at (2.25,0) {\contour{black}{$\uparrow$}};
\node[anchor=south] at (2.75,0) {\contour{black}{$\downarrow$}};
\node[anchor=south] at (3.25,0) {\contour{black}{$\uparrow$}};
\draw[black,<->] (-2.75,0.5) to[out=40,in=140] (-1.75,0.5);
\draw[black,<->] (-0.25,0.5) to[out=40,in=140] (0.75,0.5);
\draw[black,<->] (2.25,0.5) to[out=40,in=140] (3.25,0.5);
\draw[green!70!black,line width=0.3mm,-] (-2.35,0.65) to[out=-30,in=120] (-2.25,0.5);
\draw[green!70!black,line width=0.3mm,-] (-2.25,0.5) to[out=75,in=230] (-2.05,0.9);
\draw[red!80!black,line width=0.3mm,-] (0.05,0.85+0.025) -- (0.45,0.45+0.025);
\draw[red!80!black,line width=0.3mm,-] (0.05,0.45+0.025) -- (0.45,0.85+0.025);
\draw[red!80!black,line width=0.3mm,-] (0.05+2.5,0.85+0.025) -- (0.45+2.5,0.45+0.025);
\draw[red!80!black,line width=0.3mm,-] (0.05+2.5,0.45+0.025) -- (0.45+2.5,0.85+0.025);
\node[anchor=south] at (3.75,0) {$.$};
\draw[opacity=0] (0,-0.2) -- (3,-0.2);
\end{tikzpicture}
\end{align}
In a configuration in which all spins down are isolated no hopping process can occur: the aforementioned particles cannot move because they are stacked together. There are exponentially many such configurations with clustering properties, and we refer to them as jammed states.

We report a striking exception to the empirical rule that a distant localised perturbation in a low-entangled state of a quantum many-body system described by a translationally invariant Hamiltonian does not have visible effects at infinitely large times and distances. To the best of our knowledge, only one exception has been reported so far and it concerns systems prepared in the ground state when a discrete symmetry is spontaneously broken~\cite{eisler2020front,gruber2021entanglement,zauner2015time,eisler2016universal,eisler2018hydrodynamical}. 
Similarly, we consider an initial state that breaks the conserved charges that completely characterise a basis of energy eigenstates, but, differently, our initial state belongs to an \emph{exponentially} large degenerate sector. We will show that, in the limit of large time, the jammed sector is asymptotically stable under the wing-flap protocol. Quite exceptionally, we also demonstrate that the measurement results in a macroscopic change of the spin profiles on ballistic scales, namely its effect does not fade away at large times, the expectation values of local observables instead approaching nontrivial functions of the ratio between distance and time.

\paragraph{Quantum jamming.} Our discussion is specialised to the sector spanned by states $\ket{\Phi}$ satisfying the {\em jamming condition}
\be\label{eq:criterion}
\mathcal{P}_{\downarrow\downarrow}(\ell)\equiv\bra{\Phi}\frac{\1-\sigma_\ell^z}{2}\frac{\1-\sigma_{\ell+1}^z}{2}\ket{\Phi}=0,\,\,\,\forall \ell\, .
\ee
Due to~\eqref{eq:process_pictures}, such states are clearly eigenstates of $H$ with zero energy and are {\em jammed} (see also Refs~\cite{folded:1,Menon1997Conservation,HSF}). Jammed states are special since they can break the two-site shift invariance of the complete set of charges exhibited in Ref.~\cite{folded:1}. This is  possible because the spectrum has huge degeneracies associated with hidden symmetries, which have been shown, for example, to play a key role in the quench dynamics within the non-interacting sectors of the model~\cite{F:nonabelian}.

The simplest basis of the jammed sector consists of product states that are eigenstates of $\{\sigma_\ell^z\}$, but one can easily construct states with any entanglement entropy density up to $\frac{1}{2}\log 2$. For example, if $o$ denotes the sublattice of odd sites and $e$ the one of even sites, the state $\ket{\uparrow\ldots\uparrow}_o\otimes\ket{\psi}_e$ is jammed and the entanglement entropy of a spin block is half of that in the state $\ket{\psi}$. Particularly interesting are $(2n)$-site shift invariant jammed product states that are not eigenstates of $\{\sigma_\ell^z\}$, e.g.,
\be
\ket{\mathcal{L}_n}=e^{i \frac{\pi}{2n}\sum_j j\sigma_{j}^z}\ket{\uparrow \leftarrow \dots \uparrow\leftarrow}\, .\label{eq:init_state1}
\ee
This family of stationary states breaks the $U(1)$ symmetry generated by $\sum_\ell\sigma_{\ell}^z$ and, for $n>1$, also the 2-site shift invariance of the model's charges. Such states belong to the fully interacting sector of the model, which is characterised by the presence of spins up on both even and odd positions~\cite{folded:1}.  

The jammed sector is invariant under measurements of operators that commute with all elements of the set $\{\sigma_\ell^z\}$.
Note instead that measuring other observables generally results in leaving the sector.

\paragraph{Locally quasi-jammed states.}

Criterion~\eqref{eq:criterion} can be extended to inhomogeneous states that are jammed asymptotically in some scaling limit, for example,
\be\label{eq:LQJS}
\lim_{\ell,t\to\infty\atop \zeta=\ell/t \text{ fixed}} \braket{\Phi_{t}|\frac{\1-\sigma_{\ell}^z}{2}\frac{\1-\sigma_{\ell+1}^z}{2}|\Phi_{t}}=0, \qquad \forall \zeta\, .
\ee
We call them locally quasi-jammed states (LQJS).

\begin{figure}[t]
\centering
\includegraphics[width=0.49\textwidth]{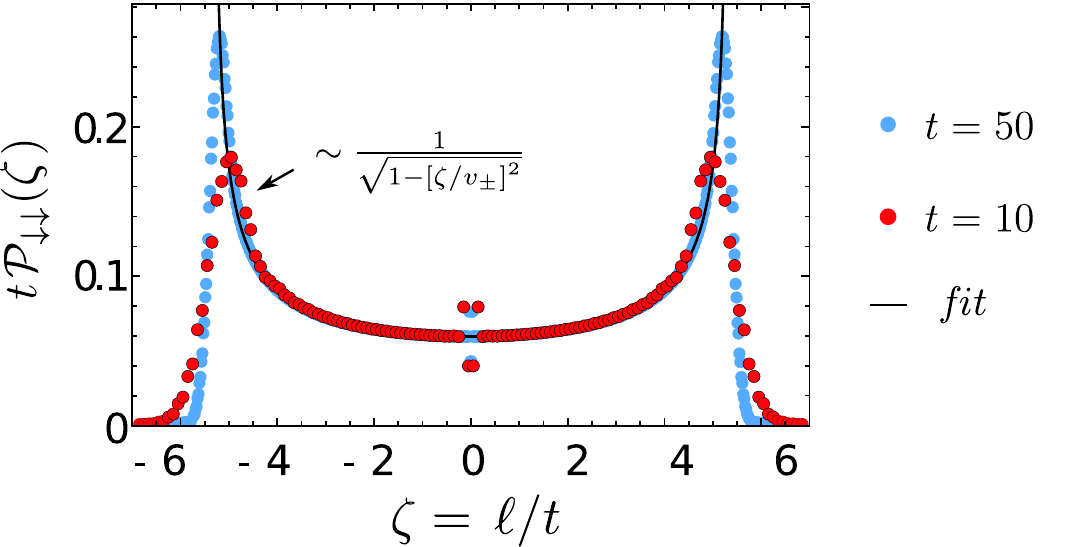}
\caption{Profile of the rescaled jamming criterion $t\,\mathcal{P}_{\downarrow\downarrow}(\ell/t)$ at times $t=10, 50$ after the flip of the central spin up (in an odd position) of the state $\ket{\mathcal{L}_1} = \ket{\uparrow \leftarrow \uparrow \leftarrow \ldots}$. The data inside the light cone are in excellent agreement with conjecture~\eqref{eq:scaling_predictions}. 
The jamming condition is locally restored as $\mathcal{P}_{\downarrow\downarrow}(\ell/t) \sim 1/t$.} 
\label{fig:Jamming_L1}
\end{figure}

We will focus on initial states belonging to the family $\ket{\mathcal L_n}$ and, for the sake of simplicity, assume 
Alice to measure $\cos(\frac{j\pi}{n})\sigma_{j}^x-\sin(\frac{j\pi}{n})\sigma^y_j$, with $j$ even, so that $\ket{\mathcal{L}_n}$ is not affected by the measurement. 
The state after Alice's measurement is then of the form 
$\ket{\Phi}=\ket{\uparrow\ldots\uparrow}_o\otimes \ket{\varphi_0}_e$. 
Let then Bob be in an odd position $2\ell'-1$ and perform a blind measurement of the spin. Since the spin is up before the measurement, the density matrix after the measurement, $\rho(0)$, will commute with $\sigma^z_{2\ell'-1}$; in particular it reads 
\be\label{eq:init_state2}
\rho(0)=\frac{2}{3}\ket{\Phi}\bra{\Phi}+\frac{1}{3}\sigma_{2\ell'-1}^x\ket{\Phi}\bra{\Phi}\sigma_{2\ell'-1}^x\, .
\ee 
The coefficients $2/3$ and $1/3$ come from considering a (blind) projective measurement -- see the Supplementary material (SM)~\cite{supplemental}. Nevertheless, the structure of the density matrix would remain the same also under more general measurement protocols. 
A local measurement can break condition~\eqref{eq:criterion} only locally and, because the defect is moved over the jammed state as a quasiparticle would be moved under the effect of a hopping Hamiltonian (in some nontrivial background -- see also Ref.~\cite{simple2021}), it is  reasonable to expect the validity of
\eqref{eq:criterion} in the limit of large time. We can then foresee an LQJS to emerge, i.e., a state where~\eqref{eq:LQJS} holds. This expectation is supported by our numerical investigation -- see Fig.~\ref{fig:Jamming_L1}.

\begin{figure}[ht!]
\centering
\includegraphics[width=0.49\textwidth]{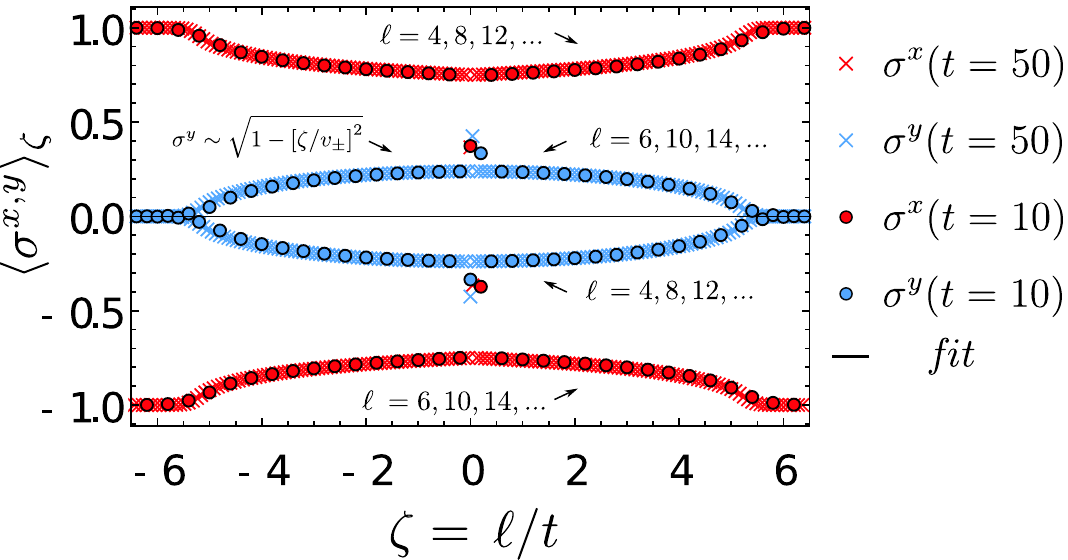}
\caption{Spin profiles $\braket{\sigma^{x}}_{\zeta}$ (red) and $\braket{\sigma^{y}}_\zeta$ (blue) 
at times  $t=10, 50$ after having flipped the central spin up (in an odd position) of the state 
$\ket{\mathcal{L}_2} = \ket{\uparrow \leftarrow \uparrow \rightarrow \ldots}$. The profile of $\sigma^{z}$ asymptotically approaches zero. Nontrivial profiles emerge on \emph{even} sites, whereas spins on odd positions relax to their original values along the $z$ direction, and are not shown. The data for  $\langle\sigma^y\rangle_{\zeta}$ are in excellent agreement with conjecture~\eqref{eq:scaling_predictions}.} 
\label{fig:Spin_profiles_L2}
\end{figure}

Notwithstanding the initial state being shift invariant by a finite number of sites, 
Bob's measurement locally breaks that symmetry. We find that, while shift invariance is locally restored, it remains globally broken, as in the systems that can be described by generalised hydrodynamics. Contrary to the latter case, however, the two-site shift invariance, which is a symmetry of a complete set of local conservation laws, is generally not restored, even locally. Figure~\ref{fig:Spin_profiles_L2}, for instance, shows that the $xy$-plane spin profiles remain staggered on the sublattice of even sites, however long the time after the initial perturbation to the state $\ket{\mathcal{L}_2}$ is: only $4$-site shift invariance is locally restored. This property is shared with non-abelian integrable systems like the quantum XY model, where the late time behaviour is generally not captured by a macro-state  characterised by the complete set of one-site shift invariant charges~\cite{F:nonabelian}.

\paragraph{Dynamics after a local spin flip.}

Since $\ket{\Phi}$ is an eigenstate, the time evolution of \eqref{eq:init_state2} 
can be immediately deduced from that of 
\be
\ket{\Psi(0)}=\sigma_{2\ell'-1}^x\ket{\Phi}\equiv (\sigma^x_{0}\ket{\uparrow\ldots\uparrow})^{}_o\otimes\ket{\varphi_0}_e\, ,
\ee
where, without loss of generality, we have set $\ell'=0$. The state at time $t$ can then be represented as follows
\be
\ket{\Psi(t)}=\sum_n(\sigma^x_n\ket{\uparrow\ldots\uparrow})^{}_o\otimes
\ket{\varphi_n(t)}_e\, ,\label{eq:invariant_subspace}
\ee
where the unnormalized wave functions $\ket{\varphi_n(t)}_e$ of spins on the sublattice of even sites satisfy
\begin{multline}
i\partial_{Jt}\ket{\varphi_{n}(t)}=K_{n-1,n}\ket{\varphi_{n}(t)}+\\
+(\mathbbm{1}-\sigma^z_{ n-1})\ket{\varphi_{n-1}(t)}+(\mathbbm{1}-\sigma^z_{ n})\ket{\varphi_{n+1}(t)}\, \label{eq:dynamical_equation}
\end{multline}
with $\ket{\varphi_{n}(0)}=\delta_{n,0}\ket{\varphi_0}$.
Here $K_{n-1,n}=\sigma_{n-1}^x\sigma_{n}^x+\sigma_{n-1}^y\sigma_{n}^y$. Note that this is a system of $\log_2 \sqrt{D}$ equations for $\ket{\varphi_n(t)}$, which are states belonging to subspaces of size $\sqrt{D}$, where $D$ denotes the size of the Hilbert space.

The dynamical equation~\eqref{eq:dynamical_equation} is block-diagonal in the Fourier space defined by
\be
\ket{\tilde \varphi_P(t)}=\sum_{n}e^{i n P}\Pi^{-n}\ket{\varphi_n(t)}\, ,\label{eq:sublattice_state}
\ee
where $\Pi$ is the $1$-site shift operator on the sublattice, such that $\Pi^{-1}=\Pi^\dag$ and $\Pi (O_\ell)_{o(e)}\Pi^\dag=(O_{\ell+1})_{o(e)}$. Specifically, we find
\be\label{eq:tildevarphi}
\ket{\tilde \varphi_P(t)}=e^{-i \tilde{H}(P) t}\ket{\varphi_0}\, ,
\ee
where the  independent time evolutions labelled by $P$ are generated by
\be
\tilde{H}(P)=J\left[K_{-1, 0}+(e^{-i P}\Pi(\mathbbm{1}-\sigma^z_{-1})+h.c.) \right]\, .\label{eq:effective_H}
\ee
Since $\Pi$ is a sublattice shift, $P$ is the eigenvalue of the momentum generating the two-site shifts on the full lattice. Indeed, denoting by $U$ the map $\ket{\varphi_n(t)}\mapsto \Pi\ket{\varphi_{n-1}(t)}$ we find
$U\ket{\tilde \varphi_P(t)}=e^{i P}\ket{\tilde \varphi_P(t)}$.

If the sites $-1$ and $0$ of the sublattice are both occupied by a spin up, the state is destroyed by $\tilde{H}(P)$: a quarter of the Hilbert space is a nullspace of $\tilde{H}(P)$.
The nontrivial action of $\tilde{H}(P)$ in the remaining space corresponds to moving a configuration of $N$ spins up on a lattice of $L=\log_2 \sqrt{D}$ sites with periodic boundary conditions through a two-site defect at $-1$ and $0$.
If both spins at $-1$ and $0$ are down, $\tilde{H}(P)$ acts as a right or left global shift of the spins up by one site. If instead there is a single spin up at $-1$ or $0$, either that spin is moved to the other site or all spins up are globally shifted in the opposite direction. Importantly, once all spins up have passed across the defect, the relative distances between them are the same as in the initial configuration. This key observation allows us to represent {\em each} configuration as an effective particle propagating by a hopping Hamiltonian (with localised defects that can be seen as a deformation of the space), 
on an extended lattice of $L+N-1$ sites. The mapping is described in the SM~\cite{supplemental}. We find the energies to be parametrised as
$
E_P(k)=4J\cos(k+P)
$, 
where the momentum $k$ of the effective particle and $P$ satisfy
\be
e^{i(L+N-1)k+ iN P}=1,\qquad e^{iL P}=1\, .\label{eq:quantisation_kP}
\ee
This mapping can also be exploited to compute the overlaps between the states and the matrix elements of the spin operators~\cite{supplemental}, providing therefore all the ingredients for the exact computation of the scaling profiles, such as the ones depicted in Figs~\ref{fig:Jamming_L1} and \ref{fig:Spin_profiles_L2}. 

For example, for odd $x$ (for the general expression see SM~\cite{supplemental}) the jamming condition \eqref{eq:criterion} after the local spin flip  can be written as (after the projective measurement there is instead an additional overall factor $1/3$)
\begin{align}
\begin{gathered}
{\cal P}_{\downarrow\downarrow}(x,t)
=\frac{1}{L^2}\sum_{N}\sum_{\chi^{(1)}_N,\chi^{(2)}_N} \braket{\varphi_0|\chi_N^{(1)}}\braket{\chi_N^{(2)}|\varphi_0}\times\\
\braket{\chi_N^{(1)}|\downarrow}\!\braket{\downarrow|^{}_{0}\,\chi_N^{(2)}}
e^{i\lceil \frac{x}{2}\rceil(P_1-P_2)+i t(E_{P_1}(k_1)-E_{P_2}(k_2))}\, .\label{eq:jamming_sum}
\end{gathered}
\end{align}
Here, $\chi_N=(k;P,\ldots)$ denotes the eigenstates and $\ldots$ stands for additional quantum numbers characterising the exponentially large sectors at fixed $k$ and $P$. The sum excludes the jammed states, since the latter are destroyed by the observable. It turns out that the sum over the additional quantum numbers can be carried out analytically, and one can reduce the entire expression to a finite number of sums, which will be discussed in a more technical work still in preparation~\cite{jammed_long_21}. 
We only anticipate that a thorough analysis of \eqref{eq:jamming_sum} shows that, first, the terms with $N/L\sim 1/2$ contribute the most and, second, the asymptotic behaviour is determined by some singular points of the averaged matrix elements of the observable (the projector on two neighbouring spins down). 

\paragraph{Numerical simulations.}
Our effective sublattice description of time evolution also has some numerical advantages. First, it
almost doubles the system size accessible to exact diagonalisation techniques. In particular, systems with $26$ spins can be simulated in a reasonable time with an ordinary laptop. As a matter of fact, it is also possible to perform semi-analytical calculations after having reduced expressions like \eqref{eq:jamming_sum} to finite numbers of sums. In that way it is possible to reach total system sizes of up to $\sim 60$ sites ($L\sim 30$) in few days of single-processor computational time (the numerical effort is expected to scale as $L^{5.5}-L^6$). 
Our main numerical checks have been however based on DMRG algorithms, which are much more efficient. 

\paragraph{Profiles.} 
We observe that, at large times, the numerical data are in excellent agreement with two ballistic-scale conjectures (i.e., $f(x,t)\to f(\zeta=\tfrac{x}{t})$)
\be 
\label{eq:scaling_predictions}
t\, \mathcal{P}_{\downarrow\downarrow}(\zeta)\approx\frac{a}{[1-(\tfrac{\zeta}{v_\pm})^2]^{\frac{1}{2}}},\ \ \langle\sigma^y\rangle_\zeta\approx b [1-(\tfrac{\zeta}{v_\pm})^2]^{\frac{1}{2}}\, ,
\ee
with $v_{\pm}=\pm 16 J/3$, and
whose assessment of validity is still in progress~\cite{jammed_long_21}. The constant prefactors in the scaling functions are well approximated by $a\approx 0.06$ and $b\approx 0.24$. These conjectures are consistent with the prediction for the maximal velocity of quasiparticle excitations on top of a jammed state $v_{\pm}[\ket{\Phi}]=\pm 8 J/\left(1+\tfrac{2}{L}\braket{\Phi|S^z|\Phi}\right)$,
which was derived in Ref.~\cite{folded:2}.

\paragraph{Can Alice recover the local state?} 
After a time $t$, the reduced density matrix of Alice's site, assumed to be at an odd distance $x+1$ from Bob's past blind measurement at site $-1$, reads 
\be
\rho_{\rm Alice}=(\1+\vec p_{x,t}\cdot\vec\sigma)/2\, ,
\ee
with $\vec p_{x,t}=\tfrac{1}{3}(\braket{\sigma^x}_\zeta-2\cos\tfrac{\pi x}{n},
\braket{\sigma^y}_\zeta+2\sin\tfrac{\pi x}{n},0
)$. Here $\zeta=x/t$, and the expectation values refer to the dynamics after the local spin flip -- see Fig.~\ref{fig:Spin_profiles_L2} and SM~\cite{supplemental} for details. For $\zeta$ in the light-cone $\vec p$ is slightly tilted from the initial (pre-measurement) direction, the tilt itself depending on Bob's position. As far as we can see, this makes it impossible for Alice to fully recover the information she had without Bob's measurement. 

\paragraph{What is general.}
We argue that a key role in the phenomenon is played by the fact that the movement of excitations over jammed states, i.e., the jamming-breaking impurities, is associated with a few-sites shift of a \emph{string} of jammed particles (see also Ref.~\cite{simple2021}). In this way the state can store  memory of its past. 

The effect seems to remain stable even under generic Hamiltonian perturbations preserving the jammed states. We have checked it for a perturbation of the form $V=g\sum_\ell\frac{\1-\sigma_\ell^z}{2}\frac{\1-\sigma_{\ell+1}^z}{2}K_{\ell+2,\ell+3}$. Flipping a single spin again results in macroscopic reorganisation of the spin profiles, albeit it now takes place in diffusive-scale coordinates $\ell\sim \sqrt{t}$ (see  SM~\cite{supplemental} for numerical evidence). Thus, the phenomenon could be observed also in non-integrable systems, in the appropriate scaling limit.

\paragraph{Summary.}
We addressed the question of how a local  measurement in a jammed state of an interacting quantum system affects the late-time dynamics of distant local observables. The measurement triggers a ballistic dynamics in which jamming is locally restored but the profile of observables remains irremediably affected by the perturbation. Arguably, this prevents the full recovery of locally damaged information in the wing-flap protocol, which is instead expected quite generally~\cite{yan2020recovery}. 

Our work gives rise to several questions. Firstly, the precise role of symmetry breaking for the observed phenomenon is yet to be clarified.  Our initial states always break some symmetries (e.g., $U(1)$, $\mathbbm{Z}_2$, or two-site shift invariance -- see also Ref.~\cite{simple2021}), however, a similar behaviour has been observed (after the completion of this paper) in a different setting where no symmetry seems to be broken~\cite{fagotti2021global}.

The second question is that of repeated projective measurements, recently identified as the cause of dynamical phase transition in certain quantum many-body systems~\cite{Li2018,Chan2019,Skinner2019,Minato2021}. Their effects on the observables, as well as on the restoration of jamming pose an intriguing open problem that could be investigated in the dual folded XXZ model even analytically.

\paragraph{Acknowledgements.}
We thank Viktor Eisler for useful discussions. This work was supported by the European Research Council under
the Starting Grant No. 805252 LoCoMacro.

During the peer-review process of this manuscript, Ref.~\cite{fagotti2021global} provided a new example of a setting where a localised perturbation remains relevant at late times.

\bibliography{jamming.bib}

\clearpage


\definecolor{naval}{HTML}{0570b0}
\definecolor{lightnaval}{HTML}{74a9cf}
\definecolor{lightred}{HTML}{e31a1c}

\onecolumngrid

\begin{center}
        {\bfseries {\em Supplementary material for}\\ ``Macroscopic effects of localised measurements in jammed states of quantum spin chains''}     
    \end{center}
    
    \begin{center}
        Kemal Bidzhiev$^1$, Maurizio Fagotti$^1$, Lenart Zadnik$^1$\\
        $^1${\em Universit\'e Paris-Saclay, CNRS, LPTMS, 91405, Orsay, France}
    \end{center}

\renewcommand{\theequation}{S.\arabic{equation}}
\renewcommand{\thefigure}{S\arabic{figure}}

\setcounter{equation}{0}
\setcounter{figure}{0}

\twocolumngrid

\subsection{The effective dynamics generated by $\tilde{H}(P)$}

Here we describe the idea behind the diagonalisation of the effective Hamiltonian $\tilde{H}(P)$ from Eq.~(12) of the main text. Detailed method will be presented in a separate publication~\cite{jammed_long_21}. Consider a configuration of $N$ spins up; the effective Hamiltonian
\be
\begin{gathered}
\tilde{H}(P)=J K_{-1, 0}+Je^{-i P}\Pi(\mathbbm{1}-\sigma^z_{-1})+\\
+Je^{i P}\Pi^{-1}(\mathbbm{1}-\sigma^z_{0})\label{eq:HtildeP}
\end{gathered}
\ee
maps it into a superposition of identical configurations shifted by one site to the right (left), with a phase $e^{-i P}$ (resp. $e^{i P}$). Repeated action of $\tilde{H}(P)$ propagates configuration in this manner until a spin up comes onto position $-1$ (or $0$). The term $J K_{-1,0}$ then kicks in: the spin up jumps to position $0$ (resp. $-1$) without acquiring a phase, while the rest of spins up retain their positions. Here is an example of nontrivial microscopic dynamics considering only the part of a superposition that moves in one direction:
\begin{center}
\hspace{-0.75cm}
\begin{tikzpicture}[scale=1]
\draw[black,->] (-2,1.25) -- (-2,-2.25) node[anchor=east] {$t$};
\draw[black,dotted] (-1.5,1.5) node[anchor=south]{\footnotesize{-$5$}} -- (-1.5,-2.5);
\draw[black,dotted] (-1,1.5) node[anchor=south]{\footnotesize{-$4$}} -- (-1,-2.5);
\draw[black,dotted] (-0.5,1.5) node[anchor=south]{\footnotesize{-$3$}} -- (-0.5,-2.5);
\draw[black,dotted] (0,1.5) node[anchor=south]{\footnotesize{-$2$}} -- (0,-2.5);
\draw[red,thick,dotted] (0.5,1.5) node[anchor=south,black]{\footnotesize{-$1$}} -- (0.5,-2.5);
\draw[red,thick,dotted] (1,1.5) node[anchor=south,black]{\footnotesize{$0$}} -- (1,-2.5);
\draw[black,dotted] (1.5,1.5) node[anchor=south]{\footnotesize{$1$}} -- (1.5,-2.5);
\draw[black,dotted] (2,1.5) node[anchor=south]{\footnotesize{$2$}} -- (2,-2.5);
\draw[black,dotted] (2.5,1.5) node[anchor=south]{\footnotesize{$3$}} -- (2.5,-2.5);
\draw[black,dotted] (3,1.5) node[anchor=south]{\footnotesize{$4$}} -- (3,-2.5);
\draw[black,dotted] (3.5,1.5) node[anchor=south]{\footnotesize{$5$}} -- (3.5,-2.5);
\node[anchor=center] at (3,1.25) {\contour{black}{$\uparrow$}};
\node[anchor=center] at (2,1.25) {\contour{black}{$\uparrow$}};
\node[anchor=center] at (1.5,1.25) {\contour{black}{$\uparrow$}};

\node[anchor=center] at (2.5,0.75) {\contour{black}{$\uparrow$}};
\node[anchor=center] at (1.5,0.75) {\contour{black}{$\uparrow$}};
\node[anchor=center,lightred] at (1,0.75) {\contour{lightred}{$\uparrow$}};

\node[anchor=center] at (2.5,0.25) {\contour{black}{$\uparrow$}};
\node[anchor=center] at (1.5,0.25) {\contour{black}{$\uparrow$}};
\node[anchor=center,lightred] at (0.5,0.25) {\contour{lightred}{$\uparrow$}};

\node[anchor=center] at (2,-0.25) {\contour{black}{$\uparrow$}};
\node[anchor=center,lightred] at (1,-0.25) {\contour{lightred}{$\uparrow$}};
\node[anchor=center] at (0,-0.25) {\contour{black}{$\uparrow$}};

\node[anchor=center] at (2,-0.75) {\contour{black}{$\uparrow$}};
\node[anchor=center,lightred] at (0.5,-0.75) {\contour{lightred}{$\uparrow$}};
\node[anchor=center] at (0,-0.75) {\contour{black}{$\uparrow$}};

\node[anchor=center] at (1.5,-1.25) {\contour{black}{$\uparrow$}};
\node[anchor=center] at (0,-1.25) {\contour{black}{$\uparrow$}};
\node[anchor=center] at (-0.5,-1.25) {\contour{black}{$\uparrow$}};

\node[anchor=center,lightred] at (1,-1.75) {\contour{lightred}{$\uparrow$}};
\node[anchor=center] at (-0.5,-1.75) {\contour{black}{$\uparrow$}};
\node[anchor=center] at (-1,-1.75) {\contour{black}{$\uparrow$}};

\node[anchor=center,lightred] at (0.5,-2.25) {\contour{lightred}{$\uparrow$}};
\node[anchor=center] at (-0.5,-2.25) {\contour{black}{$\uparrow$}};
\node[anchor=center] at (-1,-2.25) {\contour{black}{$\uparrow$}};
\draw[lightnaval,opacity=0.5] (2.66667,1.25)  -- (-0.83333,-2.25);
\filldraw[ball color=naval,opacity=0.35] (2.66667,1.25) circle (2pt);
\filldraw[ball color=naval,opacity=0.35] (2.16667,0.75) circle (2pt);
\filldraw[ball color=naval,opacity=0.35] (1.66667,0.25) circle (2pt);
\filldraw[ball color=naval,opacity=0.35] (1.16667,-0.25) circle (2pt);
\filldraw[ball color=naval,opacity=0.35] (0.66667,-0.75) circle (2pt);
\filldraw[ball color=naval,opacity=0.35] (0.16667,-1.25) circle (2pt);
\filldraw[ball color=naval,opacity=0.35] (-0.33333,-1.75) circle (2pt);
\filldraw[ball color=naval,opacity=0.35] (-0.83333,-2.25) circle (2pt);
\draw[thick,lightred,-latex] (2.16667,0.75) to [out=250,in=30,looseness=1] (1.66667,0.25);
\draw[thick,lightred,-latex] (1.16667,-0.25) to [out=250,in=30,looseness=1] (0.66667,-0.75);
\node[anchor=west,lightred] at (0.91667+0.05,-0.6) {\footnotesize{$2J$}};
\draw[thick,lightred,-latex] (-0.33333,-1.75) to [out=250,in=30,looseness=1] (-0.83333,-2.25);
\draw[thick,naval,-latex] (2.66667,1.25) to [out=250,in=30,looseness=1] (2.16667,0.75);
\node[anchor=west,naval] at (2.41667+0.06,0.9) {\footnotesize{$2Je^{iP}$}};
\draw[thick,naval,-latex] (1.66667,0.25) to [out=250,in=30,looseness=1] (1.16667,-0.25);
\draw[thick,naval,-latex] (0.66667,-0.75) to [out=250,in=30,looseness=1] (0.16667,-1.25);
\draw[thick,naval,-latex] (0.16667,-1.25) to [out=250,in=30,looseness=1] (-0.33333,-1.75);
\end{tikzpicture}
\end{center}

For the coordinate of the effective particle (in blue) we have chosen   a shifted centre-of-mass position
\be
\label{eq:effective_particle}
X(\underline{\ell})=\frac{1}{N}\sum_{j=1}^N \ell_j + \frac{N-1}{2N}(N-2N_-).
\ee 
Here $\ell_j$ are positions of spins up, $N$ their total number, and $N_-$ denotes the number of cases $\ell_j<0$. The second term in~\eqref{eq:effective_particle} ensures that $X(\underline{\ell})$ changes by $1$ even if only one of the spins up jumps between the sites $-1$ and $0$, while the rest of spins up retain their positions (the corresponding hopping of the effective particle is represented by the red curved arrows in the above diagram). The system is thus described by a hopping Hamiltonian on the lattice $\Lambda=X[\underline{\ell}]+\mathbbm{Z}$ with defects accounting for the change in the coupling constants whenever a spin up jumps between the sites $-1$ and $0$ (in red). 

The defects can be removed via a unitary transformation, which leads to a hopping model with single-particle energy levels
\be 
E_P(k)=4J\cos(k+P)\, ,
\ee
where $k$ denotes the momentum of the effective particle. The eigenstates of the effective Hamiltonian~\eqref{eq:HtildeP} are now parametrised as $\ket{k;P,\underline{d}}$, where $\underline{d}$ is a collection of distances between subsequent spins up, defined so as to be preserved by the dynamics:
\be
d_{j}=\ell_{j+1}-\ell_{j}-1+\theta(\ell_{j+1}<0)-\theta(\ell_{j}<0).
\ee

For a finite sublattice size $L=\log_2 \sqrt{D}$ one needs to consider periodic boundary conditions, where $k$ and $P$ are instead quantised according to 
\be
e^{iL P}=1,\qquad e^{i(L+N-1)k+ iN P}=1.\label{eq:quantisation_conditions}
\ee
Quantisation of $P$ follows from the momentum $P$ being associated with a shift operator $\Pi$ on the sublattice. The factor $e^{iNP}$ in the quantisation condition for $k$ is a result of the unitary transformation that removes the $N$ defects and effectively adds a phase to each hopping process caused by the defect. Factor $e^{i(L+N-1)k}$ comes from the lag acquired by the configuration w.r.t. to the position of the effective particle, when the latter traverses the system. Each time a spin up jumps from $0$ to $-1$, the right-most spin up remains fixed, except if it itself jumps between these two sites. The total acquired lag is therefore $N-1$ sites, so the configuration needs $L+N-1$ steps before returning to its initial position.

\begin{figure*}[t!]
\includegraphics[width=0.99\textwidth]{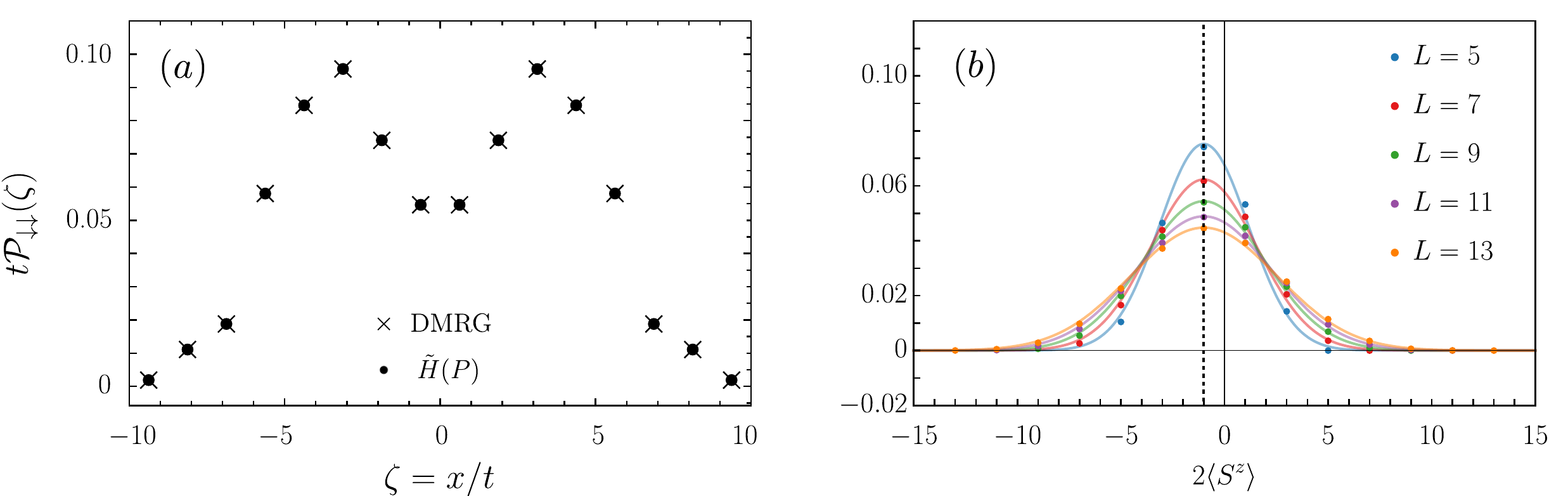}
\caption{Panel (a) shows the profile $t\mathcal{P}_{\downarrow\downarrow}(x,t)$ at $t=0.8$, for $L=13$ (a chain of $26$ spins), computed with a DMRG-based algorithm and exact diagonalisation of the effective Hamiltonians~\eqref{eq:HtildeP}. Panel (b) shows contributions to  $\mathcal{P}_{\downarrow\downarrow}$ coming from summing the elements with given $N$ (and hence $\braket{S^z}$) at $x=0$ and $t=1/2$ for the sublattice sizes $5\le L\le 13$. Data are fitted with Gaussian curves centered around the sublattice magnetisation $2\langle S^z\rangle=2N-L=-1$. The full width at half maximum scales sublinearly with the sublattice size. The term with the highest contribution is $N=(L-1)/2$.}
\label{fig:EDvsDMRG}
\end{figure*}

As a proof of concept, Fig.~\ref{fig:EDvsDMRG}(a) shows the comparison between the DMRG-based algorithm and exact-diagonalisation of Hamiltonians $\tilde{H}(P)$.

\subsection{Jamming condition $\mathcal{P}_{\downarrow\downarrow}(x,t)$}

In this section we describe the ingredients for the computation of the local jamming condition ${\cal P}_{\downarrow\downarrow}(x,t)=\tfrac{1}{4}\braket{\Psi(t)|(\mathbbm{1}-\sigma^z_x)(\mathbbm{1}-\sigma^z_{x+1})|\Psi(t)}$ in Eq.~(14) of the main text. 
Representation
\be
\ket{\Psi(t)}=\sum_n(\sigma^x_n\ket{\uparrow\ldots\uparrow})^{}_o\otimes
\ket{\varphi_n(t)}_e
\ee
of the wave function is useful to reduce the computation to the sublattice of even sites, where the condition becomes
\be
{\cal P}_{\downarrow\downarrow}=\frac{1}{2}\begin{cases} 
\braket{\varphi_{\frac{x}{2}+1}(t)|\mathbbm{1}-\sigma^z_{\frac{x}{2}}|\varphi_{\frac{x}{2}+1}(t)}&\text{$x$ even}\, ,\\
\braket{\varphi_{\frac{x+1}{2}}(t)|\mathbbm{1}-\sigma^z_{\frac{x+1}{2}}|\varphi_{\frac{x+1}{2}}(t)}&\text{$x$ odd}\, .
\end{cases}
\ee
The operators on the right-hand side act on the state of the sublattice -- their indices have been changed accordingly. Using Eqs~(10),~(11) of the main text, we then obtain
\begin{widetext} 
\begin{align}
\begin{gathered}
{\cal P}_{\downarrow\downarrow}=\frac{1}{L^2}\sum_N\sum_{\underline{d}_N^{(1)},\underline{d}_N^{(2)}}\sum_{P_1,P_2 \atop e^{iL P}=1}\sum_{k_1,k_2 \atop e^{i(L+N-1)k+ iN P}=1}e^{i\big(\frac{x}{2}+\frac{3+(-1)^x}{4}\big)(P_1-P_2)}e^{i 4Jt(\cos(k_1+P_1)-\cos(k_2+P_2))}\times\\
\times\braket{\varphi_0|k_1;P_1,\underline{d}_N^{(1)}}\braket{k_1;P_1,\underline{d}_N^{(1)}|\frac{\mathbbm{1}-\sigma^z_{-[1+(-1)^x]/2}}{2}|k_2;P_2,\underline{d}_N^{(2)}}\braket{k_2;P_2,\underline{d}_N^{(2)}|\varphi_0}\, .\label{eq:jamming_supplemental}
\end{gathered}
\end{align}
\end{widetext}

The matrix element and the overlaps between the wave functions can be computed exactly by employing the mapping between a configuration of spins up and an effective particle~\cite{jammed_long_21}. Here we only report the resulting formulas for the case $\ket{\varphi_0}=\ket{\leftarrow\ldots\leftarrow}$ (i.e. after a local measurement in the state $\ket{\mathcal{L}_1}$). For $k_1\neq k_2$ they read
\begin{widetext}
\begin{align} 
\begin{aligned}
\braket{\leftarrow\ldots\leftarrow|k;P,\underline d_N}&=
\frac{1-e^{iP}}{e^{i k}-1}\sum_{\ell=1}^N\frac{e^{i(2\ell-1-2N+\sum_{n=N+1-\ell}^{N-1}d_n)k}e^{i(\ell-1) P}}{\sqrt{2^L(L+N-1)}}\, ,\\
\braket{k_1;P_1,\underline d_N|\tfrac{\mathbbm{1}-\sigma_0^z}{2}|k_2;P_2,\underline d_N}&=
\frac{1-e^{i (P_2-P_1)}e^{i(k_2-k_1)}}{e^{i(k_2-k_1)}-1}\sum_{\ell=0}^{N-1}\frac{e^{i(2\ell+1-2N+\sum_{n=N-\ell}^{N-1}d_n)(k_2-k_1)+i\ell (P_2-P_1)}}{L+N-1}\, ,\\
\braket{k_1;P_1,\underline d_N|\tfrac{\mathbbm{1}-\sigma_{-1}^z}{2}|k_2;P_2,\underline d_N}&=e^{i(k_1-k_2)}\braket{k_1;P_1\underline d_N|\tfrac{\mathbbm{1}-\sigma_0^z}{2}|k_2;P_2,\underline d_N}\, ,
\end{aligned}
\end{align}
\end{widetext}
whereas the limit $k_1=k_2$ has to be performed carefully, taking the quantisation conditions~\eqref{eq:quantisation_conditions} into account.
We point out that the numerical computation of the fixed-$N$ terms in~\eqref{eq:jamming_supplemental} for small system sizes (via exact diagonalisation) suggests that the main contribution to the jamming condition ${\cal P}_{\downarrow\downarrow}(x,t)$ at fixed $x$ and $t$ comes from the terms with $N\sim L/2$ -- see Fig.~\ref{fig:EDvsDMRG}(b). This numerical observation can be proved analytically and holds for any sublattice size $L$~\cite{jammed_long_21}.

\subsection{Reduced density matrices}

The purpose of this section is to explain Alice's and Bob's reduced density matrices. Bob performs a blind projective measurement on one of the spins up in a factorised state. Suppose that the axis of the measurement is fixed. It can be obtained by rotating the $z$-axis by an angle $\phi\in[0,\pi)$ around the $xy$-plane unit vector $\hat{n}_\varphi=(\cos\varphi,\sin\varphi,0)$, parametrised by $\varphi\in[0,2\pi)$. Had Bob read off the result of the measurement, the spin would have collapsed into the state
\be 
\rho(\Phi,\varphi)=e^{i\tfrac{\phi}{2}\hat{n}_\varphi\cdot\vec{\sigma}}\ket{\uparrow}\bra{\uparrow}e^{-i\tfrac{\phi}{2}\hat{n}_\varphi\cdot\vec{\sigma}}\, .
\ee
In our protocol, described in the main text, the measurement is instead blind: Bob does not read off its result. Hence, due to a classical uncertainty, the state after the measurement is computed as an average w.r.t. the Haar measure:
\begin{align}
\begin{gathered}
\rho_{\rm Bob}=\int_{0}^{2\pi}\frac{{\rm d}\varphi}{2\pi}\int_{0}^{\pi}\frac{{\rm d}\phi}{\pi}\rho(\phi,\varphi)\left(\frac{\pi}{2}\cos\frac{\phi}{2}\right)=\\
=\frac{2}{3}\ket{\uparrow}\bra{\uparrow}+\frac{1}{3}\ket{\downarrow}\bra{\downarrow}\, .
\end{gathered}
\end{align}
The weight $\tfrac{\pi}{2}\cos\tfrac{\phi}{2}$ in the integral is the probability for the nonzero projection of the spin up onto the measurement axis.

At time $t$ after Bob's blind projective measurement at site $-1$,  the density matrix of the system reads
\begin{align}
\label{eq:state_supp}
    \rho(t)=\frac{2}{3}\ket{\Phi}\bra{\Phi}+\frac{1}{3}\ket{\Psi(t)}\bra{\Psi(t)},
\end{align}
where $\ket{\Psi(0)}=\sigma^x_{-1}\ket{\Phi}$, and $\ket{\Phi}=\ket{\mathcal{L}_n}$ is a jammed state given in Eq.~(4) of the main text. Alice's reduced density matrix at site $x$ (at an odd distance $x+1$ from Bob's measurement) in general reads
\begin{align}
    \rho_{\text{Alice}}=(\1+\vec p_{x,t}\cdot\vec\sigma)/2\, ,
\end{align}
where 
\begin{align}
\vec p_{x,t}=({\rm Tr}[\rho(t)\sigma^x_x],{\rm Tr}[\rho(t)\sigma^y_x],{\rm Tr}[\rho(t)\sigma^z_x])\,.
\end{align}
In the jammed state $\ket{\Phi}=\ket{\mathcal{L}_n}$ Alice finds (note that $x$ is even) 
\begin{align}
\begin{aligned}
\bra{\Phi}\sigma_x^x\ket{\Phi}&=-\cos\tfrac{\pi x}{n}\, ,\\
\bra{\Phi}\sigma_x^y\ket{\Phi}&=\sin\tfrac{\pi x}{n}\, ,\\
\bra{\Phi}\sigma_x^z\ket{\Phi}&=0\, .
\end{aligned}
\end{align}
On the other hand, in the time-evolved part of the state, $\ket{\Psi(t)}$, the expectation values of Pauli matrices asymptotically depend only on the ray $\zeta=x/t$ (we assume large distance $x$ and time $t$), i.e., $\bra{\Psi(t)}\sigma_x^\alpha\ket{\Psi(t)}=\braket{\sigma^\alpha}_{\zeta}$, $\alpha\in\{x,y,z\}$. In particular, the $z$-component of spin at an even site far from Bob's measurement is asymptotically zero, i.e., $\braket{\sigma^z}_{\zeta}=0$ (see Fig.~3 of the main text). States $\ket{\Phi}$ and $\ket{\Psi(t)}$ enter the classical mixture~\eqref{eq:state_supp} with probabilities $2/3$ and $1/3$, respectively, whence we finally obtain
\begin{align}
\vec p_{x,t}=\tfrac{1}{3}\big(\braket{\sigma^x}_\zeta-2\cos\tfrac{\pi x}{n},\braket{\sigma^y}_\zeta+2\sin\tfrac{\pi x}{n},0\big)\, .
\end{align}

\subsection{Perturbation preserving the jammed states}

\begin{figure}[h!]
\centering
\includegraphics[width=0.49\textwidth]{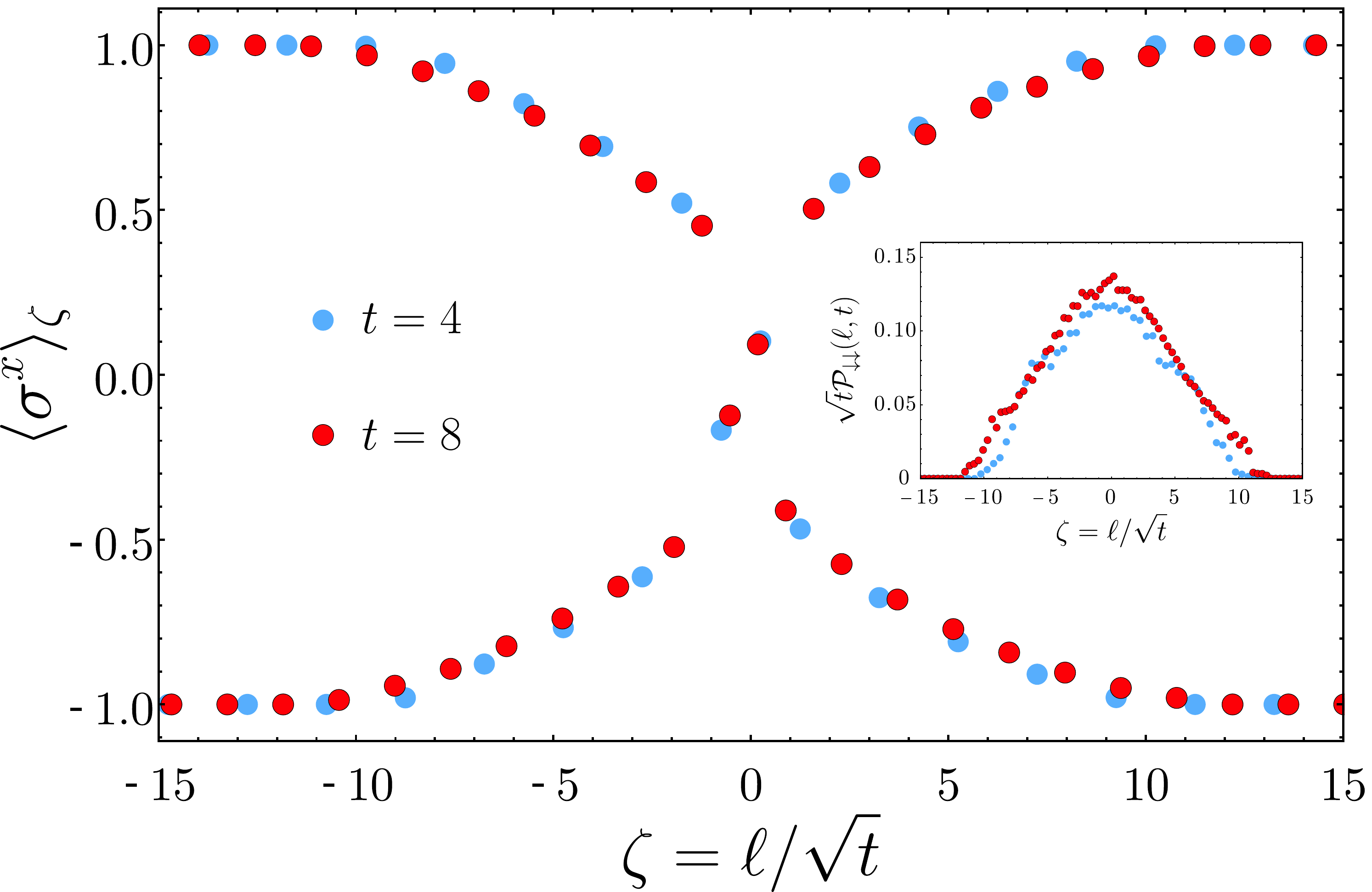}
\caption{The profile $\braket{\sigma_\ell^{x}}_t$ in the diffusive-scale coordinates $\ell/\sqrt{t}$, at times  $t=4$ (blue) and $8$ (red) after flipping up the central spin (in an odd position) of the state 
$\ket{\mathcal{L}_4} = \ket{\uparrow \leftarrow \uparrow \rightarrow \cdots}$. The initial state evolves under  $H'$ with $g=\sqrt{2}$. Nontrivial profiles emerge on \emph{even} sites, while spins on odd positions (not shown) relax to their original values along the $z$-axis. The inset shows the jamming condition $\sqrt{t}\mathcal{P}_{\downarrow\downarrow}(\ell,t)$ for different times in the diffusive-scale coordinates: $\mathcal{P}_{\downarrow\downarrow}(\ell,t)$ vanishes with time, implying the emergence of LQJS on diffusive scales. The decay of the jamming condition (up to DMRG-accessible times) seems to point at the scaling $\sim 1/\sqrt{t}$, but longer times would be required for a conclusive answer.} 
\label{fig:Spin_X_PPK}
\end{figure}

In this section we report the effects of a perturbation
that breaks integrability while preserving the jammed sector.
Specifically, we consider the Hamiltonian
\begin{align}
H' = H + g \sum_\ell\frac{\1-\sigma_\ell^z}{2}\frac{\1-\sigma_{\ell+1}^z}{2}K_{\ell+2,\ell+3}\, .
\end{align}
Fig.~\ref{fig:Spin_X_PPK} shows $\braket{\sigma^x}$ after that a spin in a jammed state of the perturbed Hamiltonian $H'$ is flipped. The expectation value exhibits diffusive scaling. This allows for a description of the local observables by macrostates depending on the diffusive-scale coordinate $\ell/\sqrt{t}$.

\end{document}